\documentclass[times, twoside, watermark]{zHenriquesLab-StyleBioRxiv}
\usepackage{blindtext}
\usepackage{hyperref}
\usepackage{tabularx}
\usepackage[numbers,sort&compress]{natbib}
\UseRawInputEncoding

\usepackage{lineno}

\leadauthor{Tokita} 

\begin{document}

\title{Networks of agglomeration: how population density rewires social networks and reshapes contagion dynamics}
\shorttitle{Population Density Networks}

\author[1]{Christopher K. Tokita}

\affil[1]{Independent Researcher, Los Angeles, CA, USA}

\maketitle

\begin{abstract}
From ancient Mesopotamia to modern cities, dense human settlements coincide with bursts of economic productivity, cultural innovation, and social change. But how does packing people more tightly together alter social organization in ways that reshape collective outcomes? Here, I use a minimal agent-based model to isolate the effect of population density, holding population size and individual behavior fixed while varying only how closely individuals are placed in space. In the model, individuals form social ties gradually, favoring those nearby and those already well-connected. Under these simple rules, varying population density alone is sufficient to reorganize social network structure: sparse populations develop locally clustered communities, while denser ones form globally integrated networks with shorter social distances and a tightly interconnected core of popular individuals. This structural transition occurs sharply over a narrow range of densities and is governed by whether physical proximity or social popularity dominates tie formation. Simulating contagions on these networks reveals that the consequences of this shift depend on what is spreading. Simple contagions (e.g., information or disease) reach a majority of individuals more quickly in denser populations. Complex contagions (e.g., social norms or collective behaviors) do not spread faster, but instead achieve broader and more reliable adoption as density increases. Together, these results show that population density can act as a structural force independent of the economic and behavioral mechanisms typically invoked to explain why cities are engines of change.
\end {abstract}


\begin{corrauthor}
\texttt{christopher.tokita{@}gmail.com}
\end{corrauthor}

\section*{Introduction}
By the 4th millennium BCE, a settlement began to expand rapidly on the banks of the Euphrates River in southern Mesopotamia. Over the course of a few centuries, it grew into Uruk, often regarded as the world’s first true city. At its peak, Uruk housed up to 50,000 residents in an area of a few square kilometers \cite{yoffee_urbanization_2015, algaze_end_2013, crusemann_uruk_2019, adams_uruk_1972}. Against a backdrop long dominated by small, kin-based villages, this concentration of people was an unprecedented shift in society. As Uruk grew, its social organization transformed. Archaeological evidence shows increased division of labor and trade specialization \cite{yoffee_urbanization_2015, algaze_uruk_1989}, the emergence of social stratification \cite{mccorriston_fiber_1997, yoffee_urbanization_2015}, and the rise of an administrative elite tied to the city's temples \cite{beale_bevelled_1978, yoffee_urbanization_2015}. Around this time, one of the world’s earliest known writing systems---proto-cuneiform---also appeared, initially as an accounting tool for navigating the increasingly complex urban economy \cite{nissen_uruk_2016, englund_accounting_2011, englund_proto-cuneiform_2004}.

In the millennia that followed, cities have repeatedly served as engines of economic prosperity, technological innovation, cultural change, and political transformation. The pattern repeats across time and cultures, from ancient Uruk and classical Athens to industrial London and modern New York City. Economists describe these phenomena as agglomeration effects: productivity, innovation, and income tend to scale superlinearly with population in urban centers \cite{moretti_new_2012, bettencourt_growth_2007, bettencourt_origins_2013}.

Humans are not unique in this regard: across the natural world, increasing the concentration of individuals coincides with shifts in social organization, although these effects are typically studied through variation in group size rather than density. Among social insects, increases in colony size are associated with greater division of labor \cite{Holbrook2011,Ulrich2018,Tokita2020,Amador-Vargas2015,Jeanson2007}, morphological specialization \cite{bourke_colony_1999}, more complex communication \cite{leonhardt_ecology_2016,donaldson-matasci_bigger_2013,Franks2006}, and enhanced collective productivity \cite{Fewell2016}, including behaviors like the fungus farming system of leaf-cutter ants \cite{swanson_welcome_2019}. Similarly, in certain species of fish, insects, birds, and mammals, large aggregations can improve vigilance against predators \cite{pulliam_advantages_1973,elgar_predator_1989,Sosna2019}, increase the efficiency of foraging and problem-solving \cite{Liker2009,Morand-Ferron2011}, and enable collective navigation as schools, swarms, and flocks \cite{Strandburg-Peshkin2015,Moussaid2009,Couzin2005}.

However, despite this evidence, it is still unclear whether these changes stem from the number of individuals or their spatial concentration. In much of the literature on collective behavior, researchers implicitly conflate population size and population density, often taking the number of individuals as a proxy for their concentration in space. Yet increasing the number of individuals in a society is not the same as increasing how densely packed they are, and few studies have explored the effect of population density in isolation. As a result, far less is known about \textit{how} population density itself shapes social organization and collective dynamics.

In this paper, I use a minimal agent-based model to test whether population density alone---without changing how many individuals there are or how any individual behaves---is enough to reorganize social networks and reshape how information and behaviors spread through them. I find that it is. Increasing population density past a critical threshold shifts networks from locally clustered to globally integrated with a core of highly connected individuals. I also find that this reorganization has distinct consequences for what spreads on them: simple contagions spread faster in denser populations, while complex contagions spread farther and take hold more reliably. Together, these results suggest that population density itself, independent of institutional or technological change, acts as a structural force that reshapes social organization and collective dynamics. It could help explain why cities, and other dense societies elsewhere in the animal world, so often are emergent centers of change, information, and innovation.

\section*{Model}

I develop a minimal agent-based model to study how population density affects the formation of social networks. In the model, individuals (nodes) occupy positions in space and form social ties (edges) with each other over time. 

In each time step, a single individual attempts to form a new social connection by probabilistically selecting a new partner based on both proximity and popularity. Individuals are more likely to connect with others who are spatially closer (proximity) and with those who already have many social connections (popularity). Aside from where they are located and how many social ties they can form, all individuals are otherwise identical.

By keeping the number of individuals in all simulations fixed and only varying the amount of space they occupy, I isolate the effect of population density. This allows us to observe how population density alone shapes social networks and the processes that unfold on them.

\subsection*{Population}

For each simulation run, I model a population of $N$ individuals living in a two-dimensional space. The space is a square with side length $d$ and open boundaries, since real-world populations do not wrap around on themselves as they would on a torus. Each individual's position is drawn independently from a uniform distribution so that, on average, individuals are spread evenly. By keeping the population $N$ fixed and varying side length $d$, I can control the population density:
$$
\delta = N / d^2
$$

\subsection*{Individuals}
All individuals begin a simulation with zero social ties and may acquire connections over time, but an individual's social capacity sets an upper limit on the total number of ties they can form. Besides their location in the space, individuals differ only in the number of social connections they are able to maintain, which I refer to as their \textit{social capacity}. 

At the start of each simulation, each individual's social capacity $c_i$ is drawn independently from a lognormal distribution with mean $\mu_c$ and coefficient of variation 0.5 (standard deviation $\sigma_c = 0.5 \mu_c$). The choice of a lognormal distribution reflects the long-tailed degree distributions observed in real-world social networks, in which most individuals maintain a modest number of relationships while a small minority sustain many more. Fixing the coefficient of variation ensures that the spread of social capacities scales proportionally with the mean.

\subsection*{Network formation}
The social network is built over a series of discrete time steps. In each time step:
\begin{enumerate}
    \item One individual $i$ is chosen uniformly at random to attempt to form a new social tie.
    \item Individual $i$ selects another individual $j$ at random based on both:
    
    \begin{enumerate}
        \item \textit{Popularity}: measured by $j$'s degree $k_j$, the total number of social ties they already have. This captures preferential attachment, a widely observed pattern in which well-connected individuals attract new connections at higher rates \cite{Barabasi1999}.
        \item \textit{Proximity}: measured by a decreasing function of the spatial distance between $i$ and $j$, denoted $F(d_{ij})$.
    \end{enumerate}
    
\end{enumerate}

Individuals that have more social ties and that are physically closer to $i$ are more likely to be chosen.

The probability that $i$ attempts to form a social tie with $j$ is:

$$
P(i \to j) = \frac{\omega_j}{\sum_{l \neq i}{\omega_l}} \;,
$$

\noindent where
$$
\omega_j = (k_j + \epsilon) F(d_{ij})
$$

\noindent is the weight assigned to individual $j$, determined both by their current degree $k_j$ and by their physical distance from individual $i$ through the function $F(d_{ij})$. The small constant $\epsilon > 0$ ensures that individuals with no existing social ties still have a non-zero probability of being selected.

$F(d_{ij})$ is a decreasing function of the distance between individuals $i$ and $j$. To model how proximity influences the likelihood that two individuals form a social tie, I use a distance-decay function common in studies on spatial interaction \cite{liben-nowell_geographic_2005, Barthelemy2011}. I define the distance weight as:

$$
F(d_{ij}) = \frac{1}{(d_{ij} + r)^\gamma}
$$

\noindent where $d_{ij}$ is the Euclidean distance between individuals $i$ and $j$, and $r$ is a distance scale that acts like a radius of local social interaction. When individuals are closer than this radius, their interaction probabilities begin to equalize, such that individuals within a small personal radius of everyday encounters (e.g., residents of the same neighborhood) are treated as similarly reachable. Without $r$, shrinking space would simply rescale all pairwise distances proportionately, leaving the relative probabilities of connection formation unchanged across different population densities. The fixed radius $r$ prevents this by establishing a spatial scale that does not change with population density.

In this structure, I follow the gravity model of social interaction, which intuitively assumes that the probability of contact declines rapidly with distance. In our model, I set $\gamma = 2$, which means that for distances large relative to $r$, doubling the distance between two individuals makes them roughly four times less likely to connect, all else being equal. This captures the tendency for most social ties to form with those nearby, while still allowing the possibility for long-distance ties. 

If the chosen pair $(i,j)$ is not already connected and both individuals have remaining capacity to add ties, an undirected edge is added between them. This process repeats until all possible ties have formed or a fixed number of time steps have elapsed, whichever comes first.

My model combines two well-established forces in social network formation. Proximity alone produces a random geometric graph, where connections depend only on distance and degree distributions are relative uniform. Real social networks, by contrast, exhibit heavy-tailed degree distributions with prominent hubs \cite{Barabasi1999, Albert2002a}. Preferential attachment is the simplest mechanism that generates this network feature, reflecting the widely documented tendency for well-connected individuals to attract new connections at higher rates \cite{Barabasi1999}. Allowing variation in social capacity further ensures that individuals differ in how many ties they can maintain, another well-observed empirical pattern \cite{dunbar_neocortex_1992, roberts_exploring_2009}. Together, these ingredients are the minimal set of assumptions needed to produce realistic networks while still isolating population density as the key variable.

\subsection*{Simulation}

\begin{table}[t]
\centering
\caption{Model parameters and baseline values used in network formation simulations.}
\label{table:parameters}
\begin{tabularx}{\columnwidth}{
  >{\hsize=0.2\hsize}X
  >{\hsize=0.5\hsize}X
  >{\hsize=0.3\hsize}X
}
\hline
Parameter & Description & Value \\
\hline
$N$ & Number of individuals & $1{,}000$ \\
$\delta$ & Population density & [$10^{-5}$, $10^5$] \\
$\mu_c$ & Mean social capacity & [5, 100] \\
$\sigma_c$ & Social capacity heterogeneity & $0.5\,\mu_c$ \\
$r$ & Interaction radius & $1$ \\
$\epsilon$ & Popularity offset & 0.1 \\
$T$ & Simulation length & $2N\mu_c$ \\
$R$ & Network replicates per parameter set & $50$ \\
\hline
\end{tabularx}
\end{table}

I implemented the agent-based model in Python using standard scientific computing libraries.

For each combination of population density $\delta$ and average social capacity $\mu_c$, I ran 50 simulations. Each simulation ran with $N=1,000$ individuals. 

To ensure that networks were given comparable opportunities to reach saturation, I scaled the length of each simulation by the average social capacity. Specifically, each simulation ran for $T = 2N\mu_c$ time steps. This meant that, on average, each individual had approximately two opportunities to form every social connection they are able to maintain. In practice, this allows networks with different parameter settings to reach an approximately stable structure, preventing us from artificially favoring higher-capacity populations.

Unless otherwise stated, all results are computed from the final network produced by each simulation run. The full set of model parameters and their values are summarized in Table \ref{table:parameters}.

\section*{Results}

\subsection*{Network Structure}

\begin{figure}[t]
\centering
\includegraphics[width=\columnwidth]{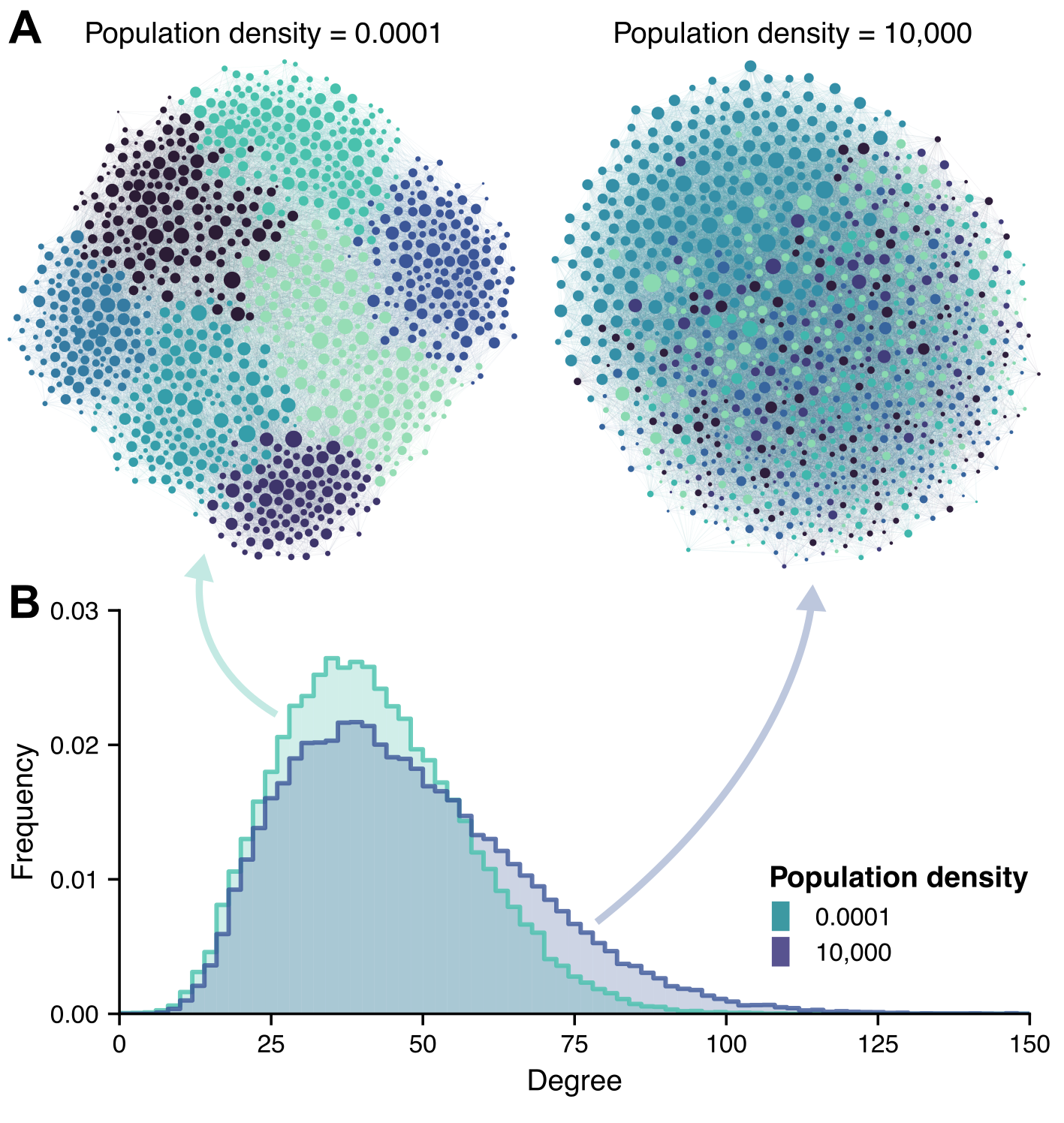}
\caption{Population density shifts networks from relatively egalitarian and clustered to more unequal and globally connected. (A) Example social network generated at low and high population density. Networks are visualized using a ForceAtlas2 layout algorithm, in which node positions reflect network topology rather than physical space. Nodes are sized by degree and colored by community membership identified by a community detection algorithm. Social networks formed at low population densities are sparser and exhibit distinct community structure of tightly connected cliques. Social networks formed at high population densities become more globally connected and unequal, with well-connected individuals forming a distinct cluster. (B) Degree distribution across 50 social network simulations at a low and high population density. All networks are simulated with a mean social capacity of $\mu_c = 50$.}
\label{fig:network_examples}
\end{figure}

Although individuals in the model follow identical behavioral rules and differ only in their social capacity, I find that increasing population density alone induces a pronounced structural transition in social networks. We can already see this shift when visually examining individual networks (Figure \ref{fig:network_examples}). At low population densities, degree distributions are relatively symmetric, and networks display clear modular structure with multiple distinct communities. At higher population densities, degree distributions develop longer right tails, indicating that a small number of individuals accumulate disproportionately more connections. The network loses its distinct local clique structure and instead features a dense core of highly connected individuals who are strongly interconnected, alongside a more diffuse periphery.

\begin{figure}[b]
\centering
\includegraphics[width=\columnwidth]{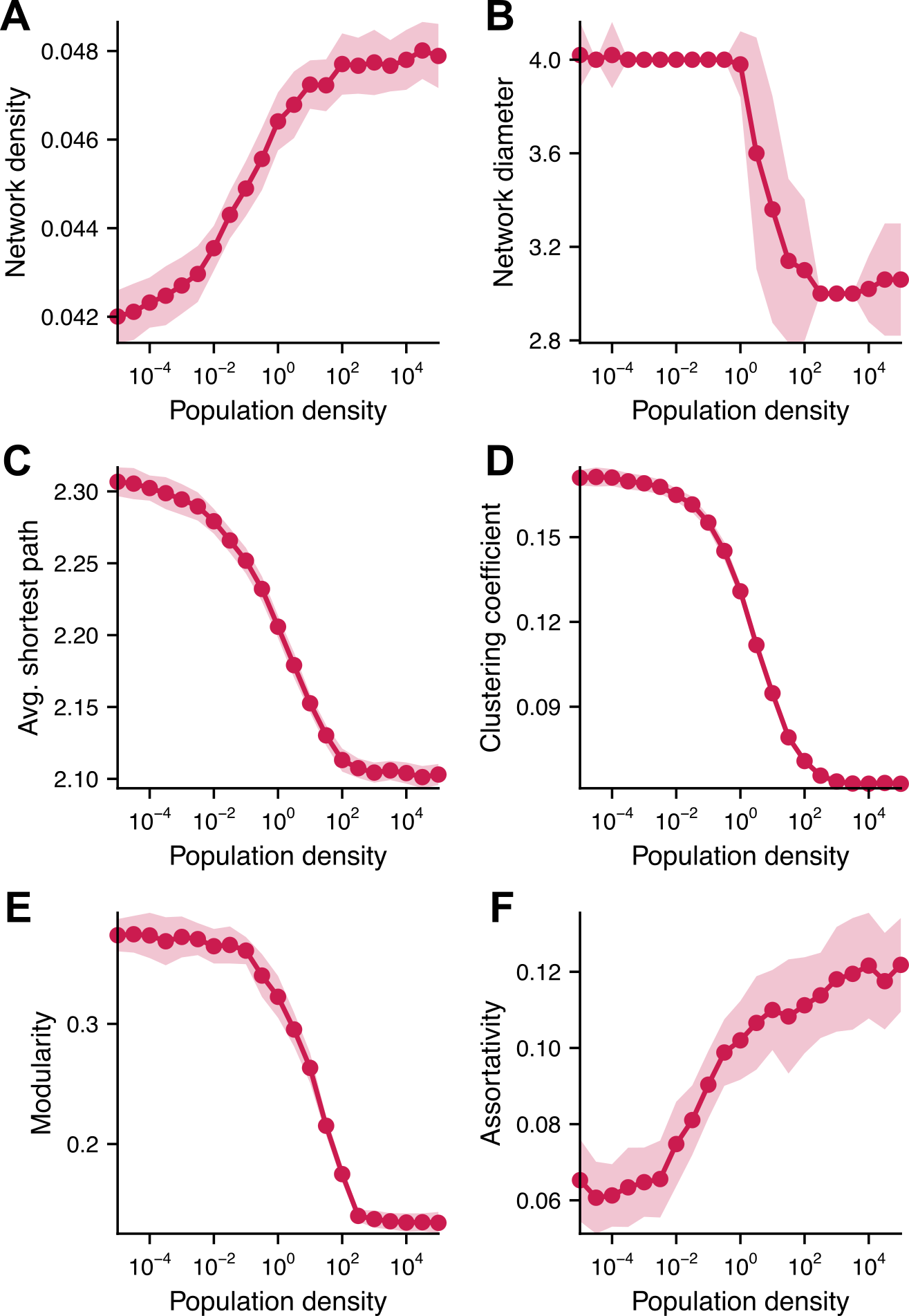}
\caption{Population density induces a transition in social network structure. Panels show how network metrics vary with population density: (A) network density, (B) network diameter, (C) average shortest path length, (D) clustering coefficient, (E) modularity, and (F) assortativity. Points represent the average ($\pm$ s.d.) of 50 social networks at each population density simulated. Social capacity is fixed at $\mu_c = 50$.}
\label{fig:density_sweep}
\end{figure}

We can quantify this transition by tracking network metrics across the full range of population densities (Figure \ref{fig:density_sweep}). At low population densities, individuals form fewer connections overall, and those connections tend to stay local among neighbors nearby. These networks have fewer ties relative to the number possible (lower average network density, A) and longer chains of intermediaries between distant individuals (larger diameters and average shortest path lengths, B,C). These networks also exhibit higher clustering and modularity (D,E), indicating that the network as a whole partitions into distinct communities where an individual's friends are also likely to be friends with one another. Assortativity is also low in this regime (F), reflecting a relatively egalitarian local structure in which highly connected and weakly connected individuals are intermixed within the same communities.

As population density increases, network structure shifts toward a more globally integrated configuration (Figure \ref{fig:density_sweep}). Networks become more densely interconnected (A), with sharply reduced diameters (B) and shorter average path lengths (C). Clustering and modularity decline as local community structure breaks down (D,E), while assortativity increases (F), reflecting the fact that highly connected individuals are increasingly likely to connect to one another. In this high population density regime, clique structure persists primarily among the most popular individuals, while the remainder of the network becomes more homogenized and spatially intermixed.

\begin{figure*}
\centering
\includegraphics[width=160mm]{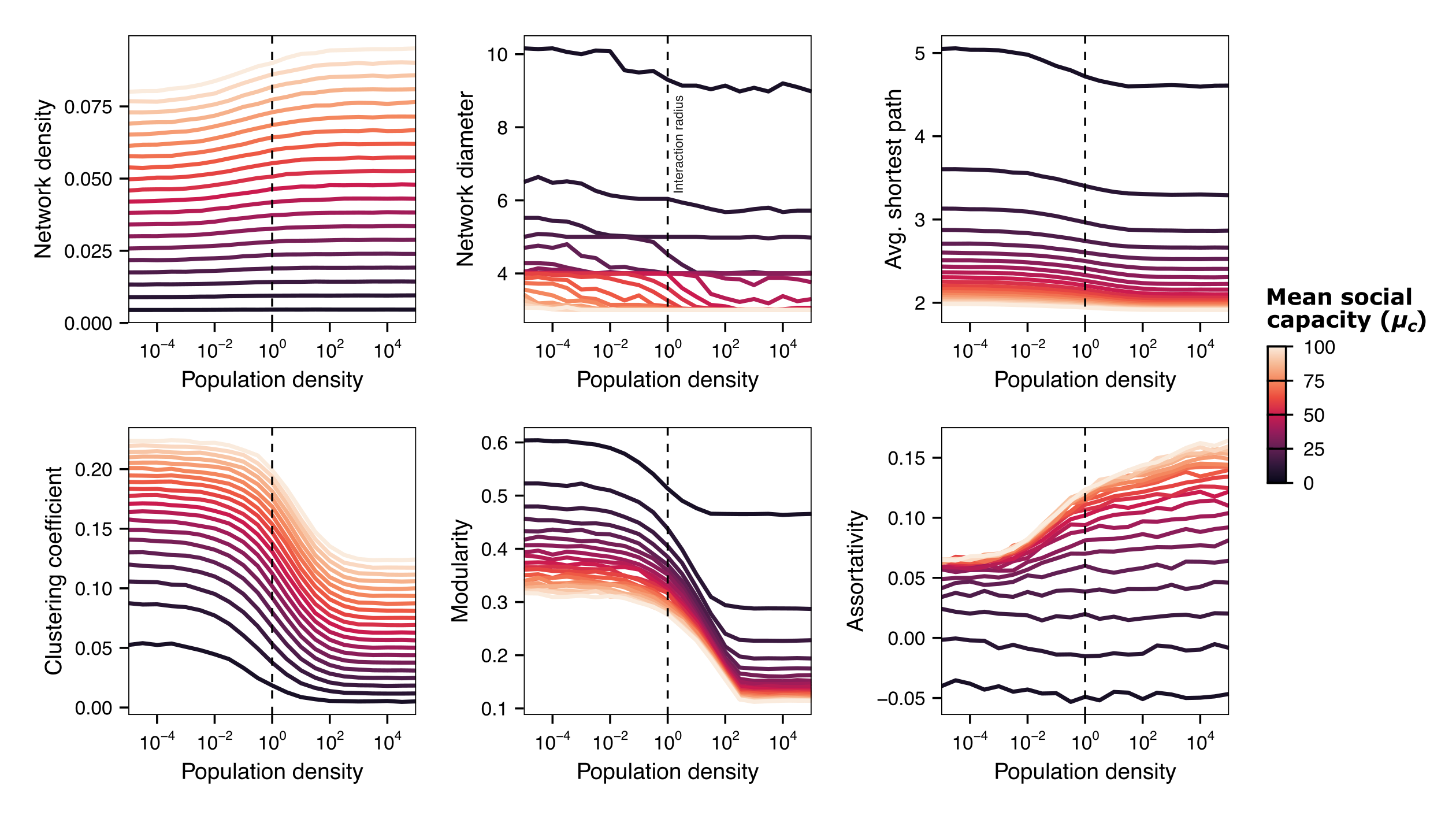}
\caption{Population density drives the structural transition in social networks, while social capacity only rescales the magnitude of connectivity. Each panel shows a network metric as a function of population density, with lines colored by mean social capacity $\mu_c$. Lines represent the average across 50 social network simulations at each combination of population density and social capacity. The dashed vertical line indicates $\delta = 1$, the density scale set by the model's interaction radius ($r = 1$).}
\label{fig:metrics_curves}
\end{figure*}

Notably, the transition in network structure occurs over a relatively narrow range of population densities and aligns closely with the model’s interaction radius. In our simulations, the transition occurs around a population density of $\delta \approx 1$, which is the scale at which physical distance begins to matter less for tie formation. Below this threshold, network growth is dominated by local, distance-driven interactions (Figure \ref{fig:proximity_vs_density}); above it, preferential attachment based on popularity increasingly governs network formation. As a result, population density shifts societies from proximity-driven, locally clustered social networks to popularity-driven, globally connected ones. Changing the interaction radius $r$---the spatial scale of everyday social encounters---shifts the population density at which the transition occurs to approximately $1 / r^2$, consistent with the prediction that networks reorganize once the typical distance between individuals matches $r$ (Figure \ref{fig:interaction_radius}). The existence and sharpness of the transition do not depend on any particular value of $r$. Instead, what matters is whether people are packed tightly enough relative to the distances over which they routinely interact.

This popularity-driven regime also reshapes who becomes well-connected. At high population density, an emergent "elite" community forms at the network's core, with popular individuals connected to one another, while less-connected individuals link both to this core and to a more intermixed remainder of the network. This elite is driven by social capacity rather than spatial position: individuals with higher social capacity accumulate disproportionately more connections, while distance from the center of the space has little predictive power (Figure \ref{fig:elite_membership}).

The results so far sweep across population density at a fixed level of social capacity ($\mu_{c} = 50$). Repeating the sweep across the full range of social capacities, I find the same transition at every level (Figure \ref{fig:metrics_curves}, \ref{fig:metrics_heatmap}). Increasing social capacity raises the average degree and reduces path lengths, but it does not fundamentally change the population density at which networks shift from locally clustered to globally integrated. Population density consistently sets where the structural transition occurs, while social capacity rescales the overall level of connectivity. Some of the metrics, such as network density, show small absolute changes because they are bounded by individual social capacity. The structural metrics most relevant to the transition, such as clustering, modularity, and path length, show much larger shifts across the population density gradient.

To further assess whether these patterns reflect smooth scaling or genuine structural transitions, I also examined network metrics by normalizing values within each level of social capacity relative to the lowest population density, $\delta = 10^{-5}$ (Figure \ref{fig:relative_heatmaps}). This normalization removes the differences in absolute magnitude across social capacities, allowing us to isolate whether the location of the structural transition shifts with social capacity. In this relative representation, the structural transition in social networks appears sharper and more regular, occurring over a narrow band of population densities that is largely independent of social capacity. Once normalized, the transition boundary is approximately vertical in density–capacity space, indicating that population density, not social capacity, determines when the network structure reorganizes.

Together, these results show that population density alone can drive a reorganization of social networks, even when individuals follow identical behavioral rules and differ only in their individual social capacity. The average social capacity of a society affects how strongly these structural changes emerge, but population density sets the threshold at which social networks shift from spatially constrained, locally clustered systems to popularity-driven, globally integrated ones.

\subsection*{Information Flow}

\begin{figure*}
\centering
\includegraphics[width=135mm]{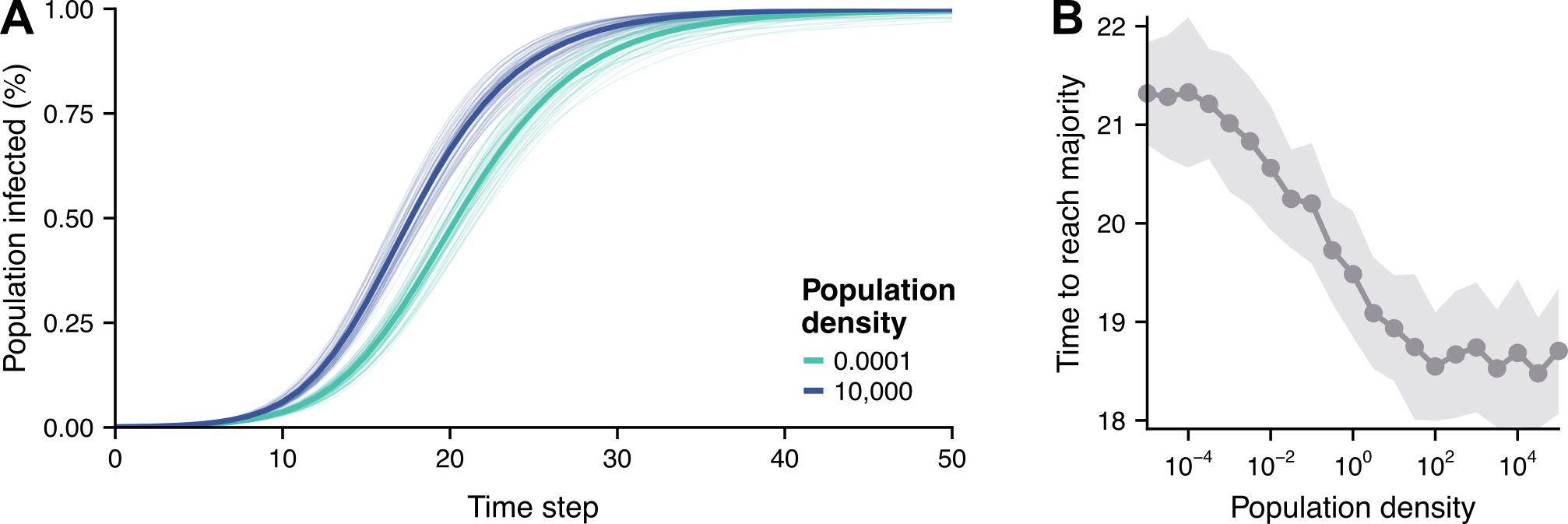}
\caption{Simple contagion dynamics on social networks formed at different population densities. (A) Simulation time series of infection at an exemplary low and high population density. Thick lines show the mean infection trajectory at each density, and thin lines show the 50 individual networks, each averaged over 50 contagion simulations. (B) The average time ($\pm$ s.e.m.) to infect a majority of individuals in a network.}
\label{fig:simple_contagion}
\end{figure*}

\begin{figure}
\centering
\includegraphics[width=\columnwidth]{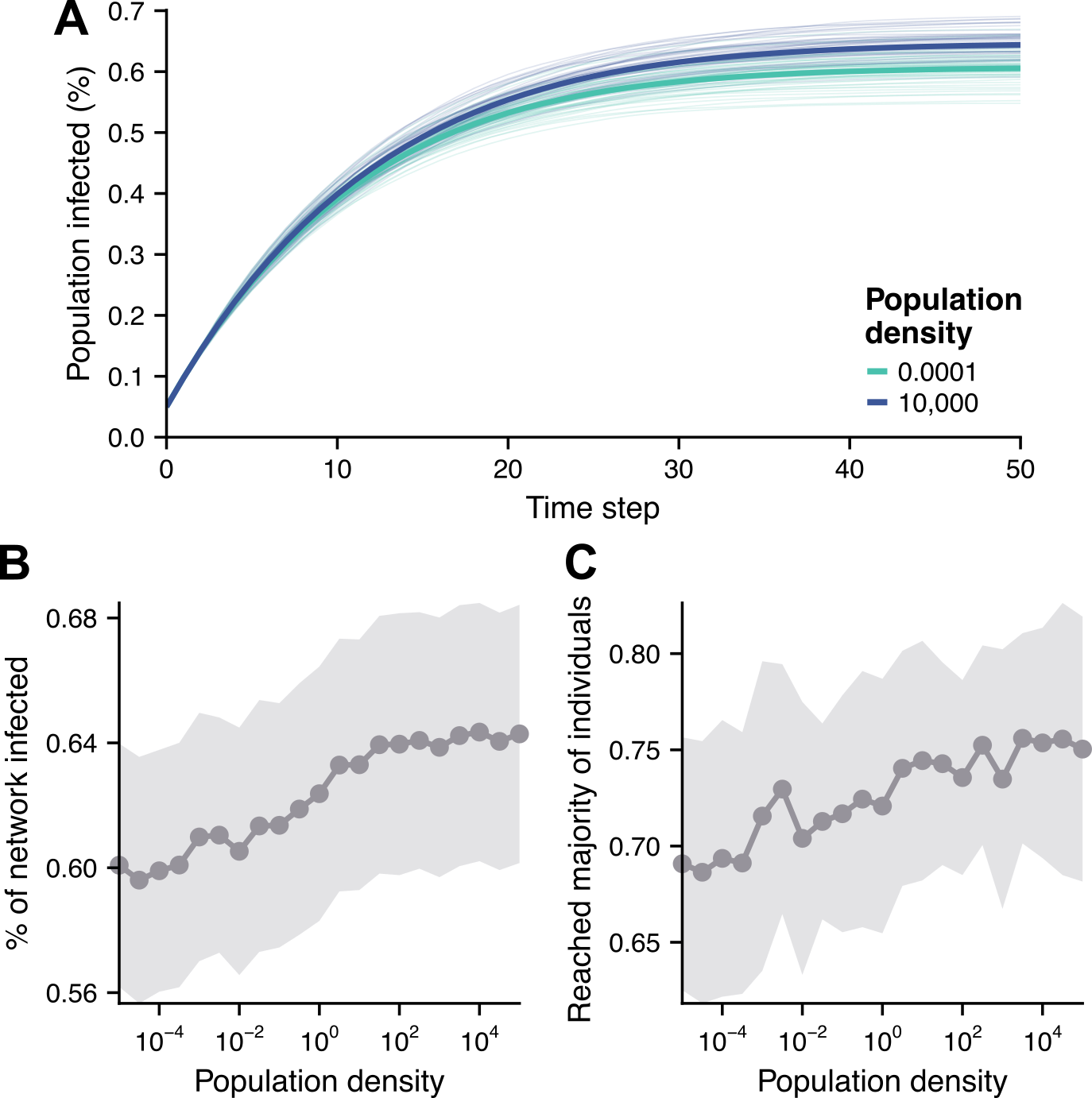}
\caption{Complex contagion dynamics on social networks formed at different population densities. (A) Simulation time series of adoption at an exemplary low and high population density. Thick lines show the mean adoption trajectory at each density, and thin lines show the 50 individual networks, each averaged over 50 contagion simulations. (B) The average percent ($\pm$ s.e.m.) of the network infected at the end of a simulation. (C) The fraction ($\pm$ s.e.m.) of contagion simulations that resulted in a majority of individuals infected.}
\label{fig:complex_contagion}
\end{figure}

To understand how population density indirectly shapes the spread of information in societies, I ran contagion simulations on the social networks generated by the network formation model. I considered two canonical classes of contagion commonly used to model spreading phenomena in social networks: simple contagions and complex contagions \cite{Centola2007, centola_how_2018}. Simple contagions capture processes such as disease transmission or the spread of news and rumors, in which exposure to a single infected individual can lead to adoption and each additional exposure independently increases the chance of infection. By contrast, complex contagions represent behaviors, social norms, or beliefs that typically require reinforcement from multiple social contacts before adoption occurs \cite{Granovetter1978, Centola2010}.

All contagions were simulated on static networks for $T = 100$ time steps. For each population density, I generated 50 independent social networks and ran 50 contagion simulations of each type on each network. I then averaged across both network realizations and contagion replicates to reduce sensitivity to stochastic initial conditions.

In the simple contagion simulations, each run began with a single infected individual chosen at random. At each time step, susceptible individuals faced a fixed per-neighbor probability of infection, $\beta = 0.01$, from each infected neighbor. Infection events were independent across neighbors, and once infected, individuals remained infected for the remainder of the simulation. For simple contagions, higher population density leads to faster diffusion (Figure \ref{fig:simple_contagion}). In all simulations, simple contagions ultimately spread to the entire population, regardless of population density. However, networks formed at higher population densities reached saturation more quickly. In particular, simple contagions reached a majority of individuals in significantly less time in high-density social networks, reflecting the shorter path lengths and greater global connectivity of high-density networks.

In the complex contagion simulations, each run began with $5\%$ of individuals infected at random. Each individual was assigned a personal adoption threshold drawn independently from a uniform distribution, $U(0,1)$, capturing how individuals in real populations differ in how readily they adopt new behaviors. Adoption thresholds are assigned independently of social capacity, so an individual's susceptibility to social reinforcement is uncorrelated with how many connections they can maintain. At each time step, a susceptible individual became infected if the fraction of their neighbors who were infected exceeded their threshold. As with simple contagions, infection was irreversible. For complex contagions, population density instead affects the extent of adoption rather than its speed (Figure \ref{fig:complex_contagion}). Unlike simple contagions, complex contagions rarely spread to the entire population, because individuals with thresholds near 1 require all of their neighbors to be infected before adopting, which makes full saturation unlikely regardless of social network structure. Population density has little effect on the speed of adoption once spreading begins. Instead, social networks formed at higher population densities exhibit wider diffusion. In these networks, complex contagions infect a larger fraction of individuals by the end of the simulation and are substantially more likely to reach majority adoption.

Taken together, these results illustrate how population density shapes diffusion processes through its effects on social network structure. Higher population density increases the global connectivity of social networks, which in turn accelerates the spread of simple contagions, while simultaneously allowing complex contagions to overcome coordination barriers and reach broader adoption. By reshaping social network structure, population density therefore also affects how social processes---such as information sharing and behavior adoption---unfold on those networks.

\section*{Discussion}
Using a simple agent-based model, I demonstrate that population density alone---even without any changes in population size, individual behavior, or social capacity---is sufficient to reshape social networks and the contagions that spread on them. As population density increases past a critical threshold tied to the scale of local social interaction, networks undergo a sharp structural transition: sparse populations develop locally clustered networks with relatively egalitarian degree distributions, while denser populations form globally integrated networks with short path lengths and an emergent core of highly connected individuals. Simulating the spread of contagions on these networks reveals that this structural shift has distinct consequences depending on what is spreading. Simple contagions spread faster in denser populations, reaching a majority of individuals in less time. Conversely, complex contagions do not spread faster but instead spread farther and take hold more reliably, resulting in broader adoption across the population. Together, these results suggest that population density acts as a fundamental structural force in societies, reshaping collective outcomes by rewiring the social connections between individuals.

The patterns of network formation and information diffusion observed in my model match empirical patterns in real cities. Research shows that social connectivity increases superlinearly with city size \cite{bettencourt_growth_2007, bettencourt_origins_2013,schlapfer_scaling_2014}, with residents in denser cities reaching each other through fewer intermediaries \cite{pan_urban_2013,schlapfer_scaling_2014}. Consistent with the model, dense urban centers develop social networks with pronounced inequality in connectivity, where a small set of highly connected individuals emerge as central hubs \cite{pan_urban_2013, herrera-yague_anatomy_2015}. These structural shifts in social networks have measurable consequences: simple contagions like disease and information spread faster in dense cities \cite{hazarie_interplay_2021, lima_understanding_2021} and innovation, which likely spreads as a complex contagion, concentrates disproportionately in larger, denser cities \cite{arvidsson_urban_2023,moretti_new_2012}. 

Conventional explanations for why cities are so productive have focused on mechanisms that emerge within already-dense economies: firms cluster geographically and gain access to specialized supply chains, thick labor markets, and knowledge spillover between workers \cite{duranton_micro-foundations_2004, rosenthal_evidence_2004, duranton_economics_2020}. Other work has instead emphasized that people behave differently when they live close together: they interact face-to-face more often, learn faster from skilled peers, and innovate at higher rates \cite{storper_buzz_2004, glaeser_agglomeration_2010}. More recently, network-based models have taken the important step of grounding these scaling patterns in social structure. For example, one model showed that urban scaling laws can emerge from the spatial interplay of social and infrastructure networks \cite{bettencourt_origins_2013}, while another model demonstrated that density-driven tie formation can create the superlinear scaling of social connectivity, productivity, and even disease rates seen in cities \cite{pan_urban_2013}. My model builds on these foundations by isolating population density from all other urban forces, holding population size and individual behavior constant. Even after stripping away these conventional factors, the model reveals a sharp structural transition in social network topology, suggesting that population density can act as a structural force independent of the economic and behavioral mechanisms typically used to explain urban agglomeration.

Population density also reshapes who benefits from increased connectivity. In denser populations, a small elite of highly connected individuals emerges. This elite arises because population density shifts the relative strength of two competing forces in tie formation: proximity and popularity-driven preferential attachment. At lower population densities, physical distance dominates, and connections form locally and relatively evenly. As density increases past the interaction radius, preferential attachment takes over, resulting in a "rich-get-richer" dynamic where already well-connected individuals attract disproportionately more connections \cite{Barabasi1999}. Empirical studies of contemporary cities reveal strikingly similar social network structure: phone-based network data from cities across multiple countries show heavy-tailed degree distributions with prominent hubs \cite{herrera-yague_anatomy_2015, pan_urban_2013}. Because social networks mediate access to resources and information, this within-city inequality in connectivity drives the superlinear scaling of wages and innovation, with larger cities sustaining larger advantage and concentrated gains to the well-connected \cite{arvidsson_urban_2023}. Archeological evidence from the world's first cities similarly shows that the first sustained periods of high population density in urban centers coincided with the emergence of social stratification and administrative elites \cite{yoffee_urbanization_2015, mccorriston_fiber_1997, ortman_settlement_2015}, despite vastly different technologies and political organization. My model offers a mechanistic explanation for why this pattern may recur across such different societies and time periods: population density systematically changes how social ties form, shifting from proximity-driven connections that produce relatively egalitarian networks to popularity-driven connections that create a connected elite.

My findings add depth to our understanding of how the structure of social networks shapes contagion dynamics. Foundational work established that network topology affects simple and complex contagions in opposing ways. Small-world networks---defined by their short global path lengths---accelerate the spread of simple contagions, like disease and information \cite{Watts1998}. But these same long-range network shortcuts can impede complex contagions, which require reinforcement from multiple overlapping contacts best found in highly clustered and modular networks \cite{Centola2007, Centola2010, Centola2007a}. This creates an apparent tradeoff: network structures that accelerate information sharing seem to impede the adoption of behaviors, and vice versa. My model complicates this picture. By showing that population density drives a sharp structural transition from highly clustered and modular networks to globally integrated ones, the model demonstrates that both types of contagion benefit from this reorganization, but just along fundamentally different dimensions. Simple contagions spread faster in the denser population regime, thanks to shorter path lengths and greater global connectivity. Complex contagions do not meaningfully speed up but instead spread farther and take hold more reliably, since the reorganized network structure allows reinforcing signals to propagate beyond local clusters. The same structural shift in a society's social network can therefore simultaneously accelerate the flow of information and disease while making collective behaviors and social norms more likely to achieve widespread adoption. My model shows that this shift can be generated upstream by the same organizing force in a society: population density.

Both types of contagion benefit from population density, but whether that helps a society depends on what spreads, and modern public health has tipped the balance toward cities. The same global connectivity that carries information and behavioral norms also carries disease, spreading innovation more widely while accelerating epidemics \cite{hazarie_interplay_2021, lima_understanding_2021, moretti_new_2012, arvidsson_urban_2023}. Dense settlements have always paid this cost: from Uruk onward, the cities that concentrated people and ideas also concentrated disease. Because these costs and benefits were coupled, for much of history cities survived by drawing migrants faster than disease could empty them \cite{de_vries_european_2013}. Modern medicine and public health have since changed this balance: innovations like sewers, water treatment, antibiotics, and vaccines have suppressed the disease cost of population density while leaving its informational and innovative benefits intact \cite{cutler_role_2005}. This may be one underappreciated reason why the modern world has urbanized so quickly. The network transformation driven by population density has remained the same, but public health now acts directly on disease transmission, blunting that cost while leaving intact the other contagion benefits that flow from the same network structure.

Although I grounded my model in the context of human societies, its mechanisms require only spatial proximity and preferential attachment, forces found across many non-human societies as well. The predictions of the model need not be restricted to humans and could be examined through experiments and observations in social insect colonies, fish schools, bird flocks, and primate groups. Across the animal kingdom, the concentration of individuals in space coincides with predictable changes in social network structure \cite{Naug2009, gagliardi_social_2023, beck_variation_2023, Tokita2020} and collective behavior, including greater division of labor \cite{Holbrook2011, Ulrich2018, Tokita2020}, more complex communication \cite{leonhardt_ecology_2016}, and enhanced collective productivity and efficiency \cite{Fewell2016, elgar_predator_1989}. However, most of this literature treats group size as the relevant variable, while population density is rarely isolated as an independent factor. My model isolates one side of this distinction by holding group size constant and varying density alone. A natural extension would vary both independently, allowing direct comparison of their relative contributions to network reorganization. Controlled experiments that manipulate population density while holding group size constant could test whether density acts as the same structural force for social organization in non-human societies as the model predicts for human ones. If so, it would point to population density as a unifying force shaping social networks and collective dynamics across societies.

Future research could extend my model by incorporating dynamic tie formation and dissolution, differences in individual behavior, or spatial mobility, but more importantly should test the current model's predictions empirically. While I have discussed how my model's predictions mirror real-world patterns, they are specific and testable predictions that warrant direct examination. A study that measures social network structure across communities spanning a range of densities and built environments---from rural villages and car-dependent suburbs to dense urban cores---could test whether clustering and modularity decline, path lengths shorten, and degree distributions develop heavier tails across a density gradient, and whether these transitions occur over a narrow band as the model predicts. The contagion predictions are also ripe for empirical testing. Evidence already shows that disease and information spread faster in denser populations.  Whether density plays the same structural role for complex contagions---enabling collective behaviors and social norms to spread farther and take hold more reliably---is a question with implications well beyond network science.

Broadly, the results of my model place population density within a larger principle: the environments we build, whether physical or digital, shape a society's collective behavior by altering how its members interact. Previous work showed that when a built environment pushes individuals towards interacting with others who are similar, as one might find in the chambers of a social insect nest \cite{Mersch2013, Pinter-Wollman2015} or the architecture of an online platform \cite{conover2011}, it can create behavioral specialization among individuals and modular communities within the social network \cite{Tokita2020}. Population density works through a different lever, the spatial scale of social interaction rather than the exact makeup of interaction partners, and can push in the opposite direction, toward a more integrated social network. Still, what unites these findings is that interaction structure imposed by an environment determines the social networks a society forms and the collective dynamics that then play out on those networks. Urbanists have long designed around this intuition, arguing that the physical form of a city---its streets, blocks, and shared gathering places---decides whether social life flourishes or thins \cite{jacobs_death_1961, alexander_pattern_1977}.

Finally, these results add a new dimension to the housing and density debates playing out in city councils, ballot initiatives, and state legislatures across the United States. Economists have established that restrictive zoning in high-productivity cities, which limits the density and amount of housing that can be built, has suppressed aggregate economic growth in the United States by 36\% over four decades \cite{hsieh_housing_2019}. My model suggests that the costs may run deeper than labor market frictions alone. Exclusionary zoning and car-dependent urban development may be holding cities on the wrong side of this transition in social organization, not just by limiting how many people live near each other, but by inflating the effective radius of daily social interaction. What matters, according to my model, is not raw population density, but how tightly packed they are relative to the spatial scale of everyday interactions. Sprawling car-dependent development can be nominally dense, while functioning more like a low-density network \cite{duranton_economics_2020, aiello_urban_2025}. Walkable mixed-use development shrinks the radius of interaction in cities. Restrictive zoning compounds the problem by creating a housing shortage that limits who can afford to live in dense neighborhoods, spatially and economically segregating the network benefits identified by my model \cite{toth_inequality_2021}. While these are speculative implications, they point to population density's role as a structural force with consequences beyond what standard economic models capture.

\medskip
\section*{Acknowledgments}
\noindent I thank Merlijn Staps and Mari Kawakatsu for their careful reading and detailed feedback on earlier drafts of this paper.

\section*{Data Availability}
All model simulation code, simulated data, and analysis code used in this paper are available on GitHub: \url{https://github.com/christokita/population-density-networks}.

\section*{References}
\renewcommand{\bibsection}{}
\setlength{\emergencystretch}{3em}
\setlength{\bibsep}{8pt}
\renewcommand\bibfont{\normalfont\sffamily\fontsize{8}{8}\selectfont}
\bibliographystyle{abbrvnat}
\bibliography{references.bib}

\newpage
\onecolumn
\captionsetup*{format=largeformat}
\section*{Supplemental Information} \label{note:SI} 
\setcounter{figure}{0}
\renewcommand{\thefigure}{S\arabic{figure}}

\begin{figure*}[ht]
\centering
\includegraphics[width=65mm]{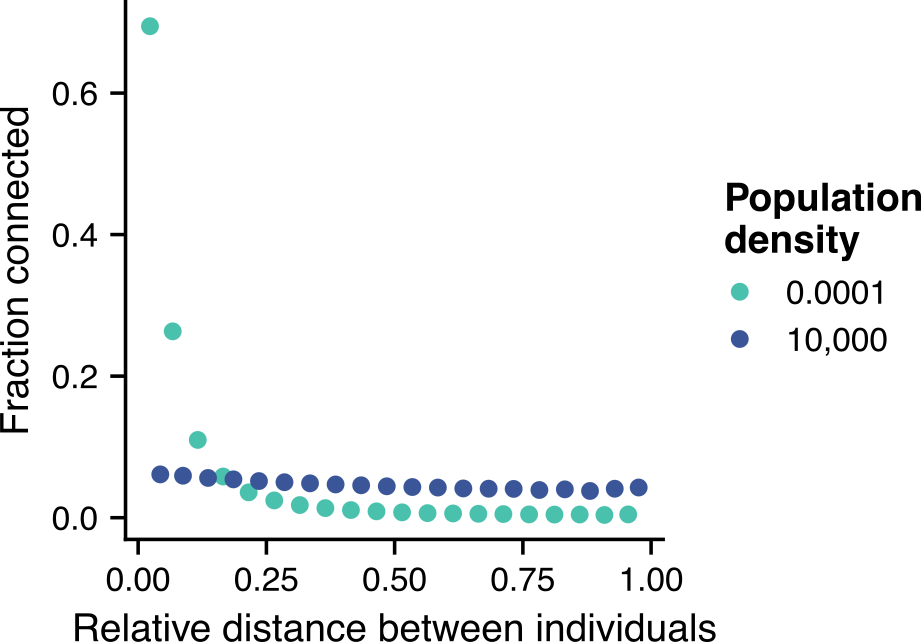}
\caption{Distance predicts social ties at low population density but not at high population density. Each point represents the fraction of sampled node pairs that are connected within a given distance bin, averaged across 5 network replicates. Distances are normalized to the maximum pairwise distance within each network. At low population density ($\delta = 0.0001$), nearby individuals are far more likely to be connected than distant ones, indicating that proximity dominates tie formation. At high population density ($\delta = 10,000$), the fraction connected is nearly uniform across distances, indicating that physical distance no longer strongly influences which individuals form ties.}
\label{fig:proximity_vs_density}
\end{figure*}

\begin{figure*}[hb]
\centering
\includegraphics[width=105mm]{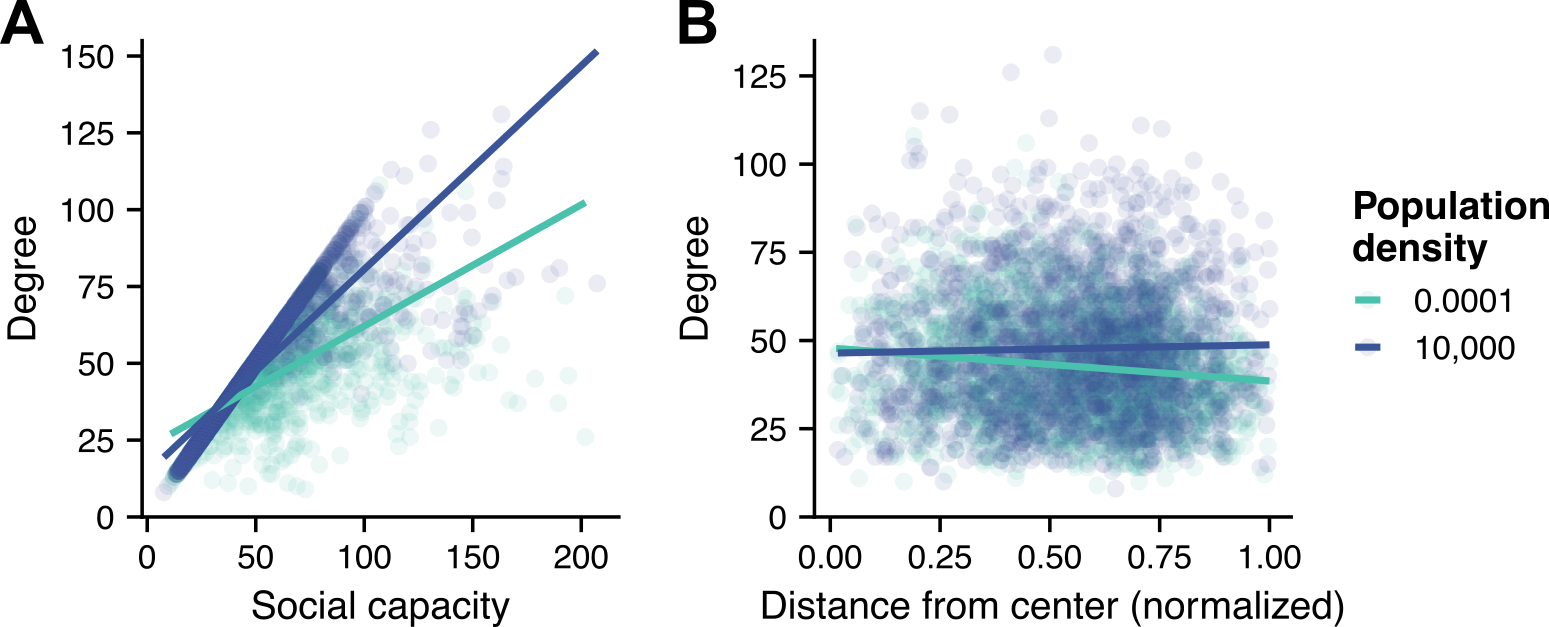}
\caption{Social capacity, not spatial location, predicts which individuals become highly connected in the network. (A) Degree as a function of social capacity at low ($\delta = 0.0001$) and high ($\delta = 10,000$) population density. Social capacity predicts degree in both regimes, but the relationship strengthens at high density---$r = 0.70$ at $\delta = 0.0001$; $r = 0.91$ at $\delta = 10,000$---reflecting the increased effect of preferential attachment. Each point represents one individual; data sampled from 50 network replicates per density.  Line represents best fit correlation. (B) Degree as a function of normalized distance from the center of the space. Spatial centrality is a weak predictor at low density ($r = 0.13$) and essentially unpredictive at high density ($r = 0.02$), indicating that the emergent elite is not an artifact of boundary effects.}
\label{fig:elite_membership}
\end{figure*}

\begin{figure}
\centering
\includegraphics[width=110mm]{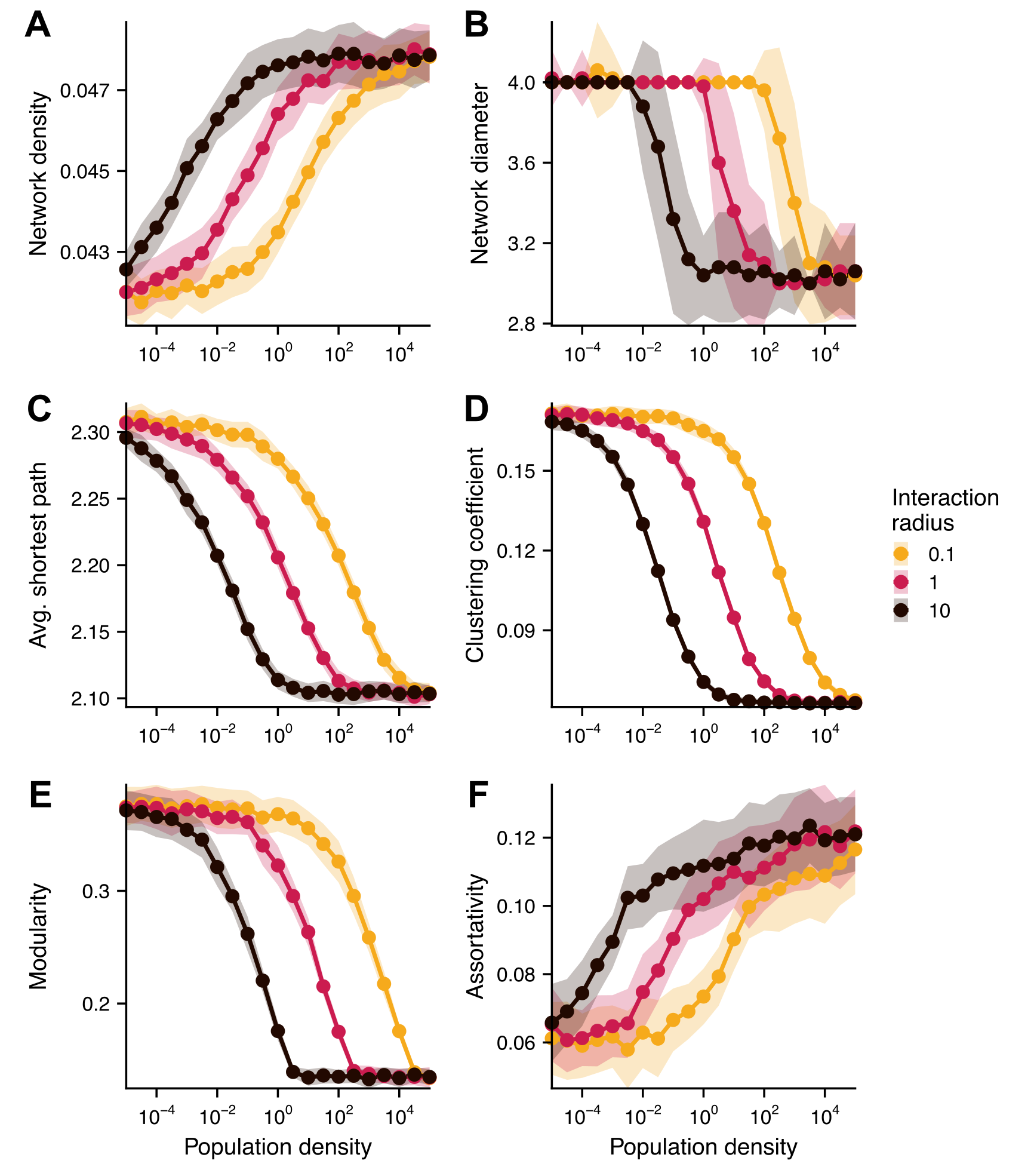}
\caption{Changing the interaction radius shifts the structural transition along population density without altering its sharpness. Panels show how network metrics vary with population density for three values of the interaction radius ($r = 0.1, 1, 10$): (A) network density, (B) network diameter, (C) average shortest path length, (D) clustering coefficient, (E) modularity, and (F) assortativity. A 10-fold decrease in $r$ shifts the transition to roughly 100-fold higher densities, consistent with the prediction that the transition occurs when typical inter-individual distance becomes comparable to $r$, i.e., the transition density scales as $1/r^2$. Points represent the average ($\pm$ s.d.) of 50 social networks at each population density. Social capacity is fixed at $\mu_c = 50$.
}
\label{fig:interaction_radius}
\end{figure}

\begin{figure*}[t]
\centering
\includegraphics[width=160mm]{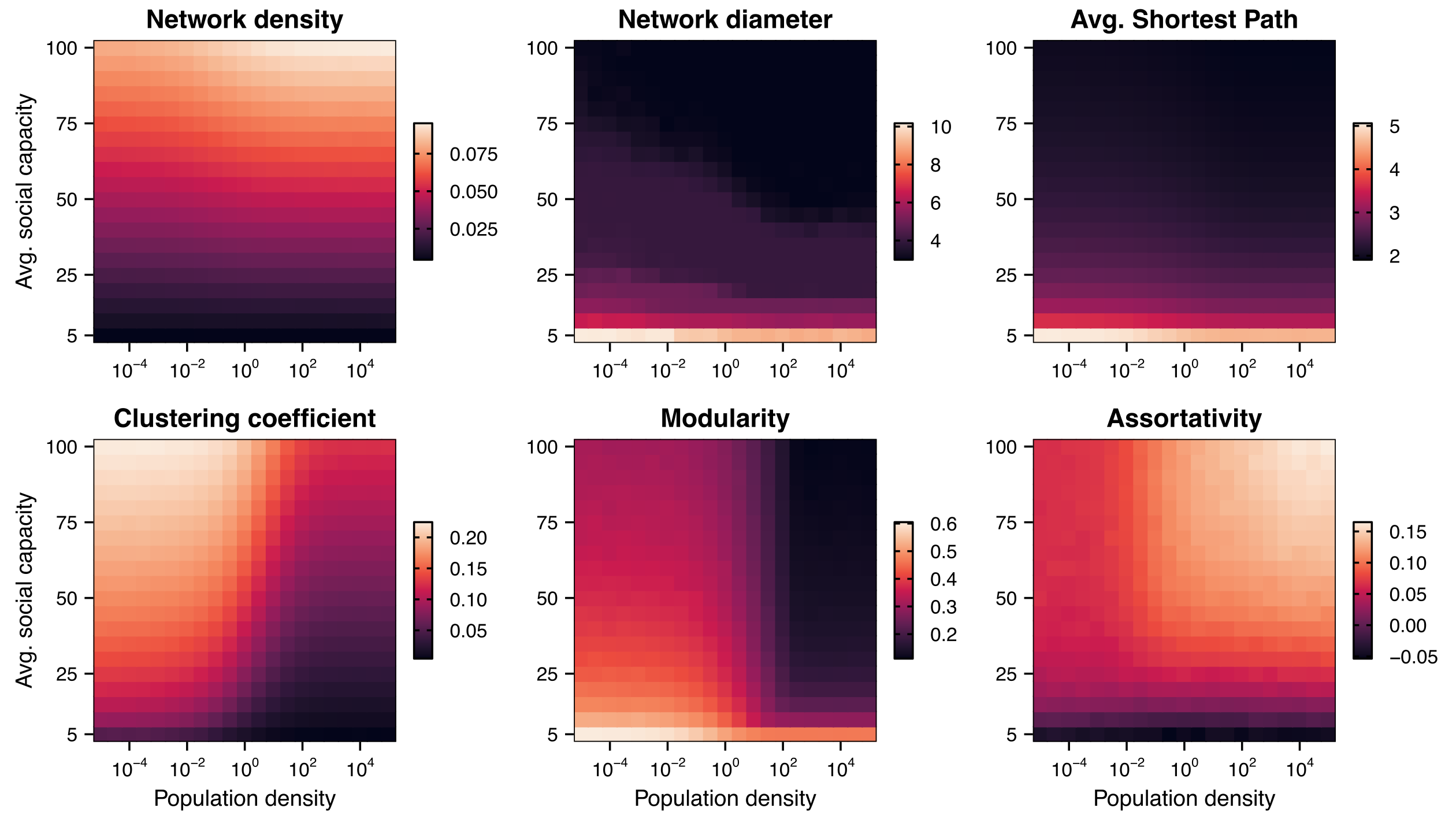}
\caption{The effect of population density and average individual social capacity on various network metrics, presented in heat map form. Each tile is colored with the average of 50 simulations.}
\label{fig:metrics_heatmap}
\end{figure*}

\begin{figure*}[ht]
\centering
\includegraphics[width=160mm]{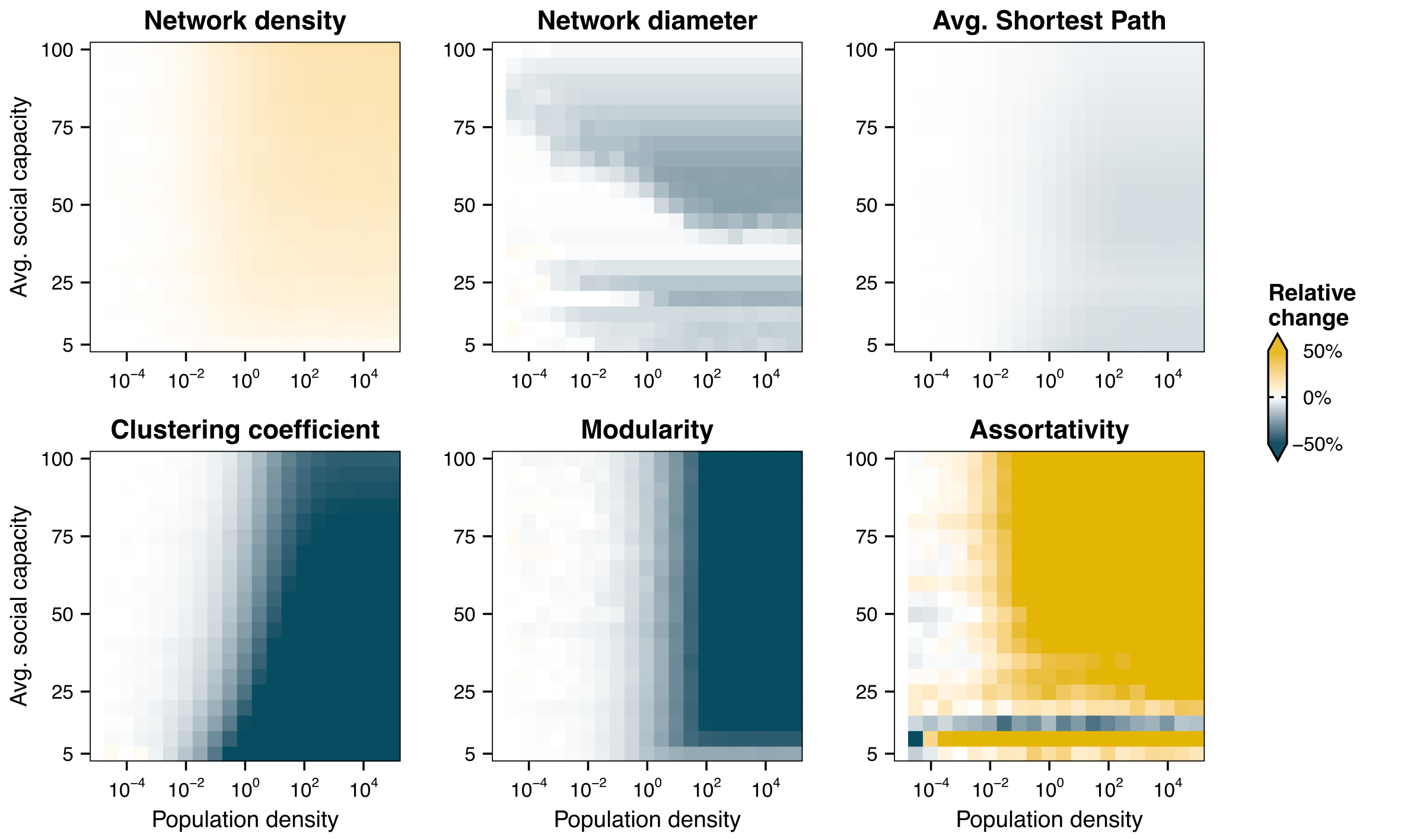}
\caption{The relative effect of population density and average individual social capacity on various network metrics. Each tile is colored with the average of 50 simulations, normalized within each level of social capacity relative to the lowest population density, $\delta = 10^{-5}$. This row-wise normalization removes magnitude differences across social capacities, isolating how population density influences the structural transition in networks.}
\label{fig:relative_heatmaps}
\end{figure*}


\end{document}